\documentclass[12pt]{article}
\usepackage{graphicx}
\usepackage{amsfonts}
\usepackage{amssymb,amsmath}
\usepackage{latexsym}
\usepackage{color}
\usepackage{cite}
\usepackage{bm}
\usepackage{ulem}

\input{colordvi.tex}

\renewcommand{\thefootnote}{\#\arabic{footnote}}

\begin{document}

\renewcommand{\thepage}{\arabic{page}}
\setcounter{page}{1}
\renewcommand{\thefootnote}{\#\arabic{footnote}}

\begin{titlepage}

\begin{center}

\vskip .5in

{\Large \bf Probing Lepton Asymmetry with 21 cm Fluctuations}

\vskip .45in

{\large
Kazunori~Kohri$\,^{1,2}$,
Yoshihiko~Oyama$\,^1$, 
Toyokazu~Sekiguchi$\,^{3,4}$ \\ and
Tomo~Takahashi$\,^5$
}

\vskip .45in

{\it
$^1$
The Graduate University for Advanced Studies (Sokendai), 1-1 Oho, Tsukuba 305-0801, Japan\\
$^2$
Institute of Particle and Nuclear Studies, KEK, 1-1 Oho, Tsukuba 305-0801, Japan\\
$^3$
Graduate School of Science, Nagoya University,
Furo-cho, Chikusa-ku, Nagoya,  464-8602, Japan\\
$^4$
Helsinki Institute of Physics, University of Helsinki, PO Box 64, FIN-00014\\
$^5$
Department of Physics, Saga University, Saga 840-8502, Japan
}

\end{center}

\vskip .4in

\begin{abstract}
We investigate the issue of how accurately we can constrain the lepton number asymmetry 
$\xi_{\nu}=\mu_{\nu}/T_{\nu}$ in the Universe 
by using future observations of 21 cm line fluctuations and cosmic microwave background (CMB).
We find that combinations of the 21 cm line and the CMB observations can constrain the lepton asymmetry 
 better  than big-bang nucleosynthesis (BBN).
Additionally, we also discuss constraints on $\xi_{\nu}$ in the presence of some extra radiation,
and show that the 21 cm line observations can substantially improve the constraints obtained by CMB alone, 
and allow us to distinguish the effects of the lepton asymmetry from the ones of extra radiation.
\end{abstract}
\end{titlepage}

\setcounter{footnote}{0}

\section{Introduction}

The issue of the asymmetry of matter and antimatter in the Universe is one of the 
important subject in cosmology and particle physics. 
The baryon asymmetry is now accurately determined by using the combination of cosmological observations such as 
cosmic microwave background (CMB), big bang nucleosynthesis (BBN), large scale structure, type Ia supernovae and so on, 
and its actual number is  
$\eta =  (n_b - n_{\bar{b}})/n_\gamma \simeq 6 \times 10^{-10}$ with $n_b, n_{\bar{b}}$ and $n_\gamma$ being 
the number densities of baryon, anti-baryon and photon, respectively.
However, on the other hand,  the asymmetry in the leptonic sector,  the lepton asymmetry,  is not well measured and
only a weak constraint for the neutrino degeneracy parameter $\xi_\nu = \mu_\nu / T_\nu$
is obtained\footnote{
  So far constraints on $\xi_\nu$ have been obtained by BBN (e.g., see
  ~\cite{Kohri:1996ke,Sato:1998nf} and Fig. \ref{fig:etaXi2014} in
  Appendix~\ref{sec:BBNrelation}), which is sometimes combined with
  CMB and/or some other observations (e.g., see
  Refs.~\cite{Popa:2008tb,Shiraishi:2009fu,Caramete:2013bua}).
}.
Although  the lepton asymmetry is expected to be the same order with the baryonic one due to the 
spharelon effect, in some models, it can be much larger than the baryonic one \cite{Casas:1997gx,MarchRussell:1999ig,McDonald:1999in,Kawasaki:2002hq,Takahashi:2003db}.
Furthermore, if  the lepton asymmetry is large, it may significantly  affect some aspects of the evolution of the Universe:
QCD phase transition \cite{Schwarz:2009ii}, large-scale cosmological magnetic field \cite{Semikoz:2009ye}, 
density fluctuations if primordial fluctuation is generated via the curvaton mechanism \cite{Gordon:2003hw,Lyth:2002my,DiValentino:2011sv} 
and so on. 

Thus it would be worth investigating to what extent the lepton asymmetry can be probed beyond the accuracy 
of current cosmological observations. Although various cosmological surveys are planned in the future, 
we in this paper consider future observations of fluctuations of neutral hydrogen 21 cm line, in addition to those of CMB,  
to study the future prospects of measuring the lepton asymmetry in the Universe.
Since the signals from the 21 cm line can cover a wide redshift range, they can be complementary to other observations such as CMB. 
In addition, the effects of the lepton asymmetry mainly appear on small scales, which can be well
measured by 21 cm observations. Thus such a survey would provide useful information. 
In this paper, to discuss 
expected constraints from the future cosmological surveys on the lepton asymmetry, or more specifically,  the degeneracy parameter $\xi_\nu$, 
we make Fisher analysis by assuming the specifications for planned observations of 21 cm fluctuations such as 
Square Kilometer Array (SKA) \cite{skaweb} and Omniscope \cite{Tegmark:2009kv}. 
We also take into account BBN and 
CMB observations such as Planck~\cite{Planck:2006aa} and CMBPol \cite{Baumann:2008aq}.

The structure of this paper is as follows. In the next section, we summarize the formulas to 
investigate the effects of the lepton asymmetry on CMB and 21 cm fluctuations. 
Then in Section~\ref{sec:results}, we present our results, paying particular attention to how 21 cm observations 
will help to probe the lepton asymmetry. Summary and conclusion of this paper is given in the final section.

\section{Lepton asymmetry}
\label{sec:formalism}

In this section, we summarize the formulas to calculate power spectra of CMB and 21 cm fluctuations
in models with non-zero lepton asymmetry, or non-zero chemical potential for neutrinos.
When there exist nonzero chemical potentials for neutrinos, they affect its energy density and pressure, 
which modifies the background evolution.   The existence of non-zero chemical potential also alters the perturbation equation. 
Below we describe the changes of the background and perturbation parts  in turn.

\subsection{Background}

The distribution function for neutrino species $\nu_i$ and its anti-particle $\bar{\nu}_i$ with $i = e, \mu, \tau$ are given by
\begin{equation}
f_{\nu_i} (p_i) =  \frac{1}{e^{p_i/T_\nu + \xi_{\nu i}} +1}, 
\qquad
f_{\bar{\nu}_i} (p_i)  =  \frac{1}{e^{p_i/T_\nu - \xi_{\nu i}} +1},
\end{equation}
where $p_i$ is momentum of $\nu_i$. $\xi_{\nu i}$ is the degeneracy parameter which is defined as $\xi_{\nu i} \equiv \mu_{\nu i} / T_{\nu}$ 
with $\mu_{\nu i}$ being the chemical potential for $\nu_i$. $T_\nu$ is the 
temperature of neutrino and related to that at the present epoch $T_{\nu_0}$ as $T_\nu = T_{\nu 0}/a$ with 
$a$ being the scale factor.

In the following,  we omit the subscript $i$ for simplicity and give the formulas
for one neutrino species including its mass $m$.
The effects of the lepton asymmetry on the background evolution appear as the changes in 
its energy density and pressure. 
The energy density and pressure of a neutrino species are given by
\begin{eqnarray}
\label{eq:rho_nu}
\rho_\nu + \rho_{\bar{\nu}} 
&=& \frac{1}{2 \pi^2} \int_0^{\infty} p^2 dp \sqrt{p^2 + m^2} 
\left( f_\nu + f_{\bar{\nu}} \right), \\
\label{eq:rho_p}
p_\nu + p_{\bar{\nu}} 
&=& \frac{1}{2 \pi^2} \int_0^{\infty} p^2 dp \frac{p^2}{3\sqrt{p^2 + m^2}}
\left( f_\nu + f_{\bar{\nu}} \right).
\end{eqnarray}
By using the comoving momentum $q \equiv pa $,  the above integral can be rewritten as 
\begin{eqnarray}
\label{eq:rho2}
\rho_\nu + \rho_{\bar{\nu}} 
&=& 
\frac{T_{\nu }^4}{2\pi^2 } \int_0^{\infty} y^3 dy \sqrt{1 + \left( \frac{a \tilde{m}}{y} \right)^2 }
\left( \frac{1}{e^{y + \xi} +1}  +  \frac{1}{e^{y - \xi} +1} \right),  \\
\label{eq:p2}
p_\nu + p_{\bar{\nu}} 
&=&
\frac{T_{\nu }^4}{6\pi^2 } \int_0^{\infty} y^3 dy \frac{1}{\sqrt{1 + \left( a \tilde{m} / y \right)^2 }}
\left( \frac{1}{e^{y + \xi} +1}  +  \frac{1}{e^{y - \xi} +1} \right),  
\end{eqnarray}
where we have defined $y$ and $\tilde{m}$ as 
\begin{equation}
y \equiv \frac{q}{T_{\nu 0}}, 
\qquad\qquad
\tilde{m} \equiv \frac{m}{T_{\nu 0 }}.
\end{equation}

Although in general, the above integrals should be performed numerically, 
in relativistic and non-relativistic limits, some useful approximation can be adopted,  in particular, 
when $ | \xi | \ll \mathcal{O}(1)$.  Below  we give explicit formulas for each case.

\bigskip
\noindent 
$\bullet$ {\bf Relativistic limit }

When $\displaystyle\frac{a \tilde{m}}{y}  (= \frac{m}{p}) \ll 1$, by expanding the integrand in Eqs.~\eqref{eq:rho_nu} and \eqref{eq:rho_p} 
up to the 2nd order in $\displaystyle\frac{a \tilde{m}}{y}$, 
the energy density and pressure can be written as 
\begin{eqnarray}
\label{eq:rho_r}
\rho_\nu + \rho_{\bar{\nu}} 
&\simeq& 
\frac{T_{\nu }^4}{2\pi^2 } \int_0^{\infty} y^3 dy \left( 1 + \frac12  \left( \frac{a \tilde{m}}{y} \right)^2 \right)
\left( \frac{1}{e^{y + \xi} +1}  +  \frac{1}{e^{y - \xi} +1} \right),  \\
p_\nu + p_{\bar{\nu}} 
&\simeq&
\frac{T_{\nu }^4}{6\pi^2 } \int_0^{\infty} y^3 dy \left( 1 - \frac12  \left( \frac{a \tilde{m}}{y} \right)^2 \right)
\left( \frac{1}{e^{y + \xi} +1}  +  \frac{1}{e^{y - \xi} +1} \right).
\end{eqnarray}
These integrals can be performed exactly and we obtain 
\begin{eqnarray}
\label{eq:rho_r2}
\rho_\nu + \rho_{\bar{\nu}} 
&\simeq& 
 \frac{7 \pi^2 }{120} T_\nu^4 \left[ 
\left\{ 1+ \frac{30}{7} \left( \frac{\xi}{\pi} \right)^2 + \frac{15}{7} \left( \frac{\xi}{\pi} \right)^4  \right\}
+ \frac{5}{7 \pi^2} (a \tilde{m})^2 \left\{ 1+ 3 \left( \frac{\xi}{\pi} \right)^2 \right\} 
\right], 
\\
p_\nu + p_{\bar{\nu}} 
&\simeq&
\frac13 \frac{7 \pi^2 }{120} T_\nu^4 \left[ 
\left\{ 1+ \frac{30}{7} \left( \frac{\xi}{\pi} \right)^2 + \frac{15}{7} \left( \frac{\xi}{\pi} \right)^4  \right\}
- \frac{5}{7 \pi^2} (a \tilde{m})^2 \left\{ 1+ 3 \left( \frac{\xi}{\pi} \right)^2 \right\} 
\right].
\end{eqnarray}

\bigskip
\noindent 
$\bullet$ {\bf Non-relativistic limit } 

When $ \displaystyle\frac{a \tilde{m} }{ y }  (= \frac{m}{p} )\gg 1$,  
we can expand Eq.~\eqref{eq:rho_nu}  around $y /(a\tilde{m}) = 0$ and $\xi = 0$  
\footnote{
In a non-relativistic limit for any $\xi$ values, 
the exact solutions of $\rho_{\nu}+\rho_{\bar{\nu}}$ and $p_{\nu}+p_{\bar{\nu}}$
are  expressed by using polylogarithm. 
The formulas are given in Appendix~\ref{sec:appB}.
}
as
\begin{eqnarray}
\rho_\nu + \rho_{\bar{\nu}} 
&=&
\frac{T_{\nu }^4}{2\pi^2 } \int_0^{\infty} y^3 dy  
 \frac{a \tilde{m}}{y}
\sqrt{\left( \frac{y}{a \tilde{m}} \right)^2  + 1}
\left( \frac{1}{e^{y + \xi} +1}  +  \frac{1}{e^{y - \xi} +1} \right) \notag \\
&\simeq  &
\frac{T_{\nu }^4 a \tilde{m}}{2\pi^2 } \int_0^{\infty} y^2 dy  
\left[ 1 + \frac12 \left( \frac{y}{a \tilde{m}} \right)^2  \right]
\left( \frac{1}{e^{y + \xi} +1}  +  \frac{1}{e^{y - \xi} +1} \right) \notag \\
&\simeq &
\frac{T_{\nu }^4 a \tilde{m}}{2\pi^2 } \int_0^{\infty} y^2 dy  
\left[ 1 + \frac12 \left( \frac{y}{a \tilde{m}} \right)^2  \right]
\sum_{i} C_i (y) \xi^i ,
\end{eqnarray} 
where $C_i (y)$ are the coefficients for the expansion of $ \left( (e^{y+\xi} +1)^{-1} + (e^{y-\xi} +1)^{-1} \right)$ around 
$\xi =  0$. 
We note that the terms with odd power for $\xi$ do not appear.
Explicit formulas for $C_i (y)$ are given in Appendix~\ref{sec:appA}.

Having the expressions for  $C_i (y)$, we can analytically perform the integral of the form:
\begin{equation}
\int_0^\infty C_i(y) y^2 dy, 
\qquad
{\rm and}
\qquad
\int_0^\infty C_i(y) y^4 dy.
\end{equation}
By taking into account the terms up to the 10th order in $\xi$, we obtain
\begin{eqnarray}
\label{eq:rho_nr}
\rho_\nu + \rho_{\bar{\nu}}  
&\simeq&
 \frac{T_{\nu }^4}{2\pi^2 } ( a \tilde{m}) 
\left[  3 \zeta (3) + (\log 4) \xi^2  + \frac{1}{24} \xi^4  -\frac{1}{1440}  \xi^6  + \frac{1}{40320}\xi^8  - \frac{17}{14515200} \xi^{10} \right] \notag \\
&& 
+  \frac{T_{\nu }^4}{4\pi^2 }\frac{1}{ a \tilde{m}}
\left[ 45 \zeta(5)   + 18 \zeta (3) \xi^2 + (\log 4) \xi^4 + \frac{1}{60} \xi^6  - \frac{1}{6720} \xi^8 + \frac{1}{302400}\xi^{10} \right],
\notag \\
\end{eqnarray}
where $\zeta(x)$ means the Riemann zeta function.
Similar calculations also hold for the pressure, and we have, up to the 10th order in $\xi$, 
\begin{eqnarray}
\label{eq:p_nr}
p_\nu + p_{\bar{\nu}} 
&\simeq&
\frac{T_{\nu }^4}{6\pi^2 } 
 \frac{1}{a \tilde{m}}
\left[ 45 \zeta(5)   + 18 \zeta (3) \xi^2 + (\log 4) \xi^4 + \frac{1}{60} \xi^6  - \frac{1}{6720} \xi^8 + \frac{1}{302400}\xi^{10} \right]
\notag \\
&- &
\frac{T_{\nu }^4 }{12\pi^2 } 
\left( \frac{1}{a \tilde{m}} \right)^3
\left[  \frac{2835 \zeta (7)}{2}  + 675 \zeta(5) \xi^2 + 45 \zeta (3) \xi^4  +  (\log 4) \xi^6  + \frac{1}{112} \xi^8  -\frac{1}{20160} \xi^{10}\right]. 
\notag \\
\end{eqnarray} 

We have checked that above formulas are accurate as $10^{-7}$ for $|\xi | < 1$ to obtain $\rho_\nu$ and $p_\nu$ 
with non-zero $\xi$. 

\subsection{Perturbation equation}

Here we discuss the perturbation equation for massive neutrinos including the chemical potential.
By perturbing the phase-space distribution function $f_\nu $ as \cite{Ma:1995ey}
\begin{equation}
\delta f_\nu (\tau, \vec{x}, \vec{p}) + \delta f_{\bar{\nu}} (\tau, \vec{x}, \vec{p}) = 
\left(\bar{f}_\nu (p)  +  \bar{f}_{\bar{\nu}} (p) \right)\Psi_\nu (\tau, \vec{x}, \vec{p}),
\end{equation}
where $\bar{f}_\nu$ and $\bar{f}_{\bar{\nu}}$ are the background distribution functions,
 and $\Psi_\nu$ represents its perturbation.
$\tau$ is the conformal time.
The perturbed Boltzmann equation for $\Psi_\nu$ for the Fourier mode $\vec k$ in the synchronous gauge is given by
\begin{equation}
\dot{\Psi}_\nu+ i \frac{y}{\sqrt{y^2 + a^2 \tilde{m}^2}} (\vec{k}\cdot \hat{n} ) \Psi_\nu 
+ \frac{d \ln (\bar{f}_\nu + \bar{f}_{\bar{\nu}} ) }{d \ln y} \left[  
\dot{\eta_{{\rm T}}} - \frac12 \left( \dot{h_{{\rm L}}} + 6 \dot{\eta}_{{\rm T}} \right) (\vec{k}\cdot \hat{n} )^2
\right] =0,
\end{equation}
where $h_{{\rm L}}$ and  $\eta_{{\rm T}}$ are metric perturbations, and a dot represents the derivative with respect to the conformal time
 (we follow the notations in \cite{Ma:1995ey}).
$\hat{n}$ is the direction of the momentum $\vec{p}$.
We expand $\Psi_\nu$ with the Legendre polynomial as 
\begin{equation}
 \Psi_\nu (\tau, \vec{k}, \vec{p}) = \sum_{l=0}^{\infty} (-i)^l (2l+1)\Psi_{\nu l} (\tau, \vec{k}, p) P_l (\hat{k}\cdot \hat{n}),
\end{equation}
with $\hat{k}$ being the direction of $\vec{k}$. The evolution equations for each multiple moment in the synchronous gauge take the form:
\begin{eqnarray}
\dot{\Psi}_{\nu 0} & = &  
- \frac{y k}{\sqrt{y^2 + a^2 \tilde{m}^2}} \Psi_{\nu 1} + \frac16 \dot{h}_{{\rm L}} \frac{ d \ln (\bar{f}_\nu + \bar{f}_{\bar\nu})}{d \ln y}, \\ \notag \\
\dot{\Psi}_{\nu 1} & = &   
\frac{y k}{3 \sqrt{y^2 + a^2 \tilde{m}^2}} \left( \Psi_{\nu 0} - 2 \Psi_{\nu 2} \right), \\ \notag \\
\dot{\Psi}_{\nu 2} & = &   
\frac{y k}{5 \sqrt{y^2 + a^2 \tilde{m}^2}} \left( 2 \Psi_{\nu 1} - 3 \Psi_{\nu 3} \right)  
-\left(  \frac{1}{15} \dot{h}_{{\rm L}}  + \frac25 \dot{\eta}_{{\rm T}} \right) \frac{ d \ln (\bar{f}_\nu + \bar{f}_{\bar\nu})}{d \ln y}, \\ \notag \\
\dot{\Psi}_{\nu l} & = &   
\frac{y k}{(2l+1) \sqrt{y^2 + a^2 \tilde{m}^2}} \left( l \Psi_{\nu (l-1)} - (l+1) \Psi_{\nu (l+1)} \right), ~~ ({\rm for} ~l \ge 3).
\end{eqnarray}
The dependence on the chemical potential appears in the factor $d \ln (\bar{f}_\nu + \bar{f}_\nu)/d \ln y$, which 
can be written as \cite{Lesgourgues:1999wu}
\begin{equation}
\frac{ d \ln (\bar{f}_\nu + \bar{f}_{\bar\nu})}{d \ln y} 
= - \frac{y \left( 1+ \cosh \xi \cosh y \right)}{(\cosh \xi + \exp(-y) ) (\cosh \xi + \cosh y)}.
\label{eq:dlnfdlny}
\end{equation} 

By making the modifications given above as well as those for the background quantities to {\tt CAMB} \cite{Lewis:1999bs}, 
we calculate power spectra of CMB and 21 cm fluctuations and make a Fisher matrix analysis, 
whose results will be discussed in the next section.

\section{Results}
\label{sec:results}

Now in this section, we discuss future prospects of  the determination of the lepton asymmetry, or the chemical 
potentials for neutrino. For this purpose, we study expected constraints on $\xi$ by making Fisher analysis 
adopting future observations of 21 cm fluctuations and CMB.  
In the analysis, we assume the specifications of SKA \cite{skaweb} and Omniscope  \cite{Tegmark:2009kv}
for 21 cm fluctuations and Planck \cite{Planck:2006aa}\footnote{
Although the temperature data from Planck is already available, 
here we are going to combine CMB data with future 21cm observations, 
and hence we also treat Planck in the same manner as other future observations.
}
and CMBPol \cite{Baumann:2008aq} for CMB.
Our methodology in the following analysis is basically the same as our previous one \cite{Kohri:2013mxa}, thus 
we refer the readers to  \cite{Kohri:2013mxa} for the details.

\begin{figure}[htbp]
  \begin{center}
    \resizebox{120mm}{!}{
     \includegraphics{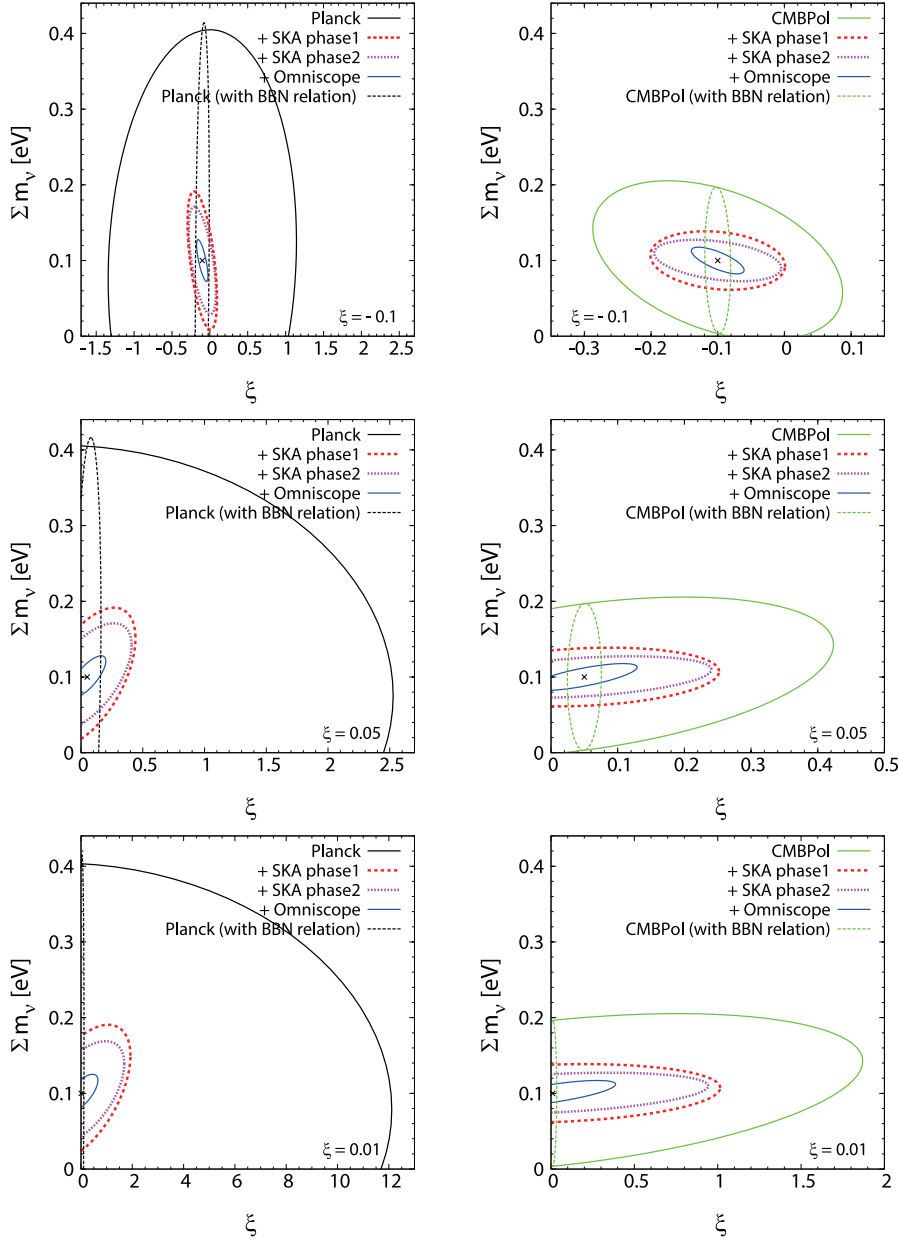}
}
\end{center}
\caption{Expected 2$\sigma$ constraints on the $\sum m_\nu$--$\xi$ plane. 
  As CMB data, the Planck and CMBPol surveys are adopted in the left and right panels, respectively.
  In order from top to bottom, the fiducial values of $\xi$ are set to $-0.1$, 0.05 and 0.01. 
  Here we mainly present constraints for fixed $Y_p=0.25$. 
  Shown are the constraints from CMB alone (solid black/green line) as well as the ones from CMB data combined with 21~cm data from 
  SKA phase1 (red line), SKA phase2 (magenta line) and Omniscope (blue line). 
  As a reference, the constraints from CMB data alone with the BBN relation
  are also shown (dotted black/green line).
  Note that scales in $x$-axis differ among different panels.
  }
  \label{fig:xi_mnu_noBBN}
\end{figure}

\begin{figure}[htbp]
  \begin{center}
    \resizebox{120mm}{!}{
     \includegraphics{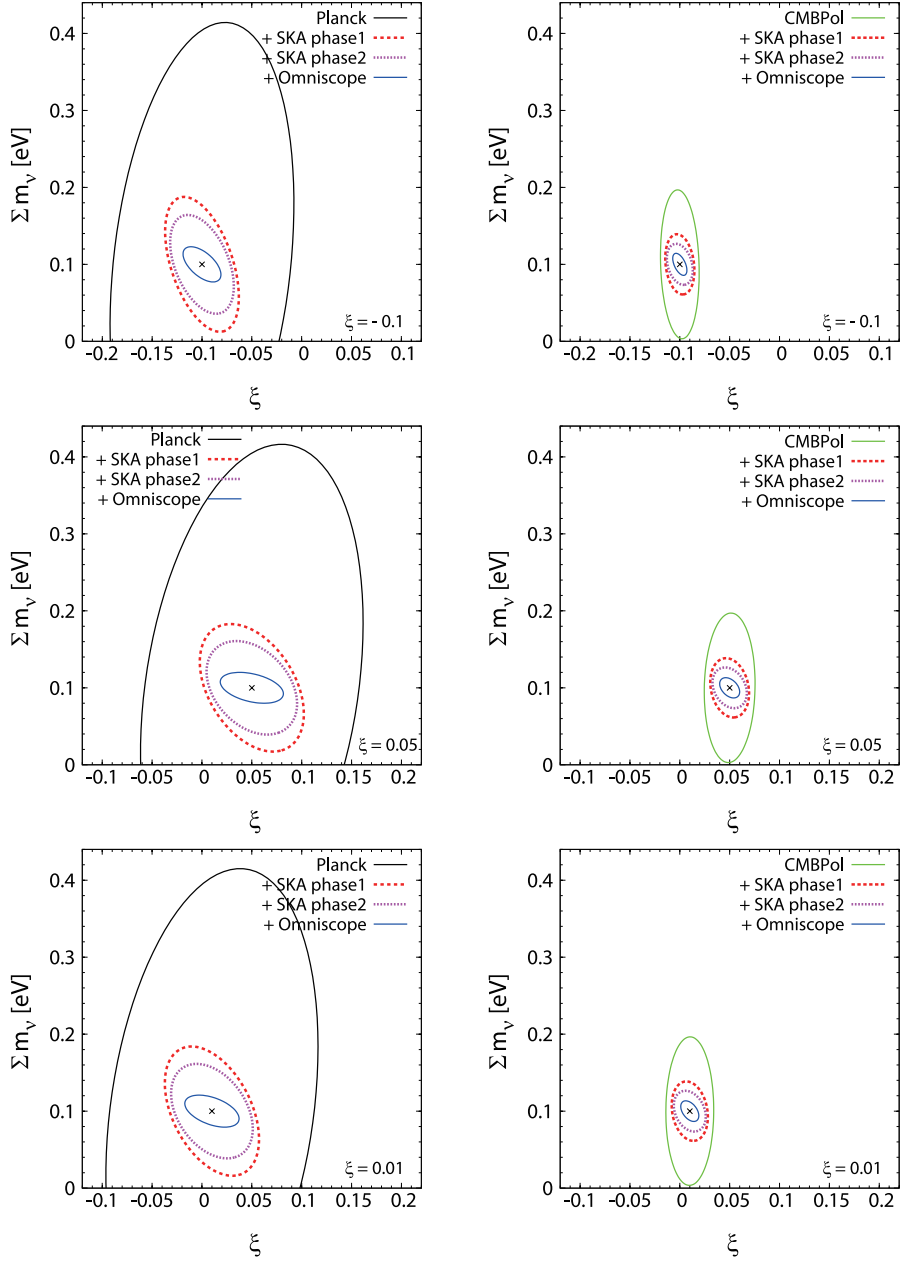}
}
  \end{center}
  \caption{Expected 2$\sigma$ constraints on the $\sum m_\nu$--$\xi$ plane. In this figure, the BBN relation is assumed.}
  \label{fig:xi_mnu_wBBN}
\end{figure}

In the following analysis, we explore the parameter space which includes 
the degenerate parameter 
$\xi=\xi_{\nu_{e}}=\xi_{\nu_{\mu}}=\xi_{\nu_{\tau}}$ assuming the universal lepton asymmetry\footnote{
Regardless of the initial value of $\xi_{\nu i}$ (with $ i= e,\mu,\tau$) at the decoupling, 
the lepton asymmetry would be universal, due to the large mixing in neutrino mass matrix~\cite{Dolgov:2002ab}.}
and neutrino mass $m_\nu$ as well as 
the six standard cosmological parameters
($\Omega_\Lambda,~\Omega_b h^2,~\Omega_mh^2,~\tau_{\rm reion},~A_s,~n_s$)\footnote{
Here $\Omega_\Lambda$, $\Omega_b$ and $\Omega_m$ are respectively energy densities of 
the cosmological constant, baryon and matter, $h$ is the Hubble parameter normalized as $H_0 = 100 h~{\rm km/s/Mpc}$, 
$\tau_{\rm reion}$ is the optical depth for reionization, $A_s$ and $n_s$ are respectively the amplitude and the spectral index for the primordial power 
spectrum.
}. In addition to these parameters, in some cases, we also include the helium abundance $Y_p$ and the effective number 
of neutrino species for extra (dark) radiation $\Delta N_\nu$
which gives its energy density in units of a single massless neutrino species as 
\begin{equation}
\bar\rho_{\rm ext}=\Delta N_\nu \frac{7\pi^2}{120}{T_\nu}^4.
\end{equation}
Although the chemical potential $\xi$ can be regarded as the changes to 
$N_\nu$, that is, the effective number of 
neutrino species for total dark radiation (neutrinos and extra radiation) as seen from 
Eqs.~\eqref{eq:rho_r2} and \eqref{eq:rho_nr}, $\Delta N_\nu$ counts for possible other contribution to $N_\nu$.
Furthermore, in BBN theory, $Y_p$ is related to
$\Omega_bh^2, \xi$  and 
$\Delta N_\nu$.
Therefore we make the analysis with/without assuming so-called BBN relation among these parameters in some analysis.  
When the BBN relation is not adopted, we vary $Y_p$ freely or fix it to $Y_p=0.25$. 

Regarding fiducial parameters, we often present constraints for several fiducial values of $\xi$ and $\Delta N_\nu$.
On the other hand, fiducial values of $\sum m_\nu$ is fixed  to be 0.1~eV and 
those of other cosmological parameters are fixed to be
($\Omega_\Lambda,~\Omega_b h^2,~\Omega_mh^2,~\tau_{\rm reion},~A_s,~n_s$)
$=( 0.6914, 0.02216, 0.1417, 0.0952, 2.214\times10^{-9}, 0.9611 )$,
which are the best fit values from the Planck result \cite{Ade:2013zuv}.

\subsection{Cases without extra radiation}
Let us first see the cases without extra radiation.
Fig.~\ref{fig:xi_mnu_noBBN} shows constraints on the $\xi$--$\sum
m_\nu$ plane for mainly the cases where we fixed $Y_p$ to 0.25 without
assuming the BBN relation.  On the other hand, constraints only from
CMB observations with the BBN relation $Y_p(\Omega_bh^2,~\xi,~\Delta
N_\nu)$ are also shown as well, for the purpose of comparison.
Regarding fiducial values of $\xi$, we adopted $\xi=0.01$, 0.05 and
$-0.1$ here. Note that $\xi=0.05$ and -0.1 roughly correspond to the
upper and lower bounds at 2$\sigma$ from primordial abundance of the
light elements (See Fig.~\ref{fig:etaXi2014} in
  Appendix~\ref{sec:BBNrelation} and
Ref. \cite{Steigman:2012ve}).
From the figure, we can immediately see that 21~cm observations can be a powerful probes of the lepton asymmetry.
Compared with the constraints on $\xi$ from Planck alone, 
the error is improved by a factor around 5 (10) by combining SKA (Omniscope).
Even though CMBPol can by itself give much tighter constraints than Planck, combinations with 21~cm observations are still 
able to improve the constraints further by a factor around 2 (SKA) and 4 (Omniscope).
We also note that constraints on the neutrino masses from CMB observations can 
be also improved by combining 21~cm observations.
As an illustrative example, constraints on cosmological parameters for the cases with fiducial $\xi=0.05$
are summarized in Table \ref{tab:fixedYp}. 

In Fig.~\ref{fig:xi_mnu_noBBN}, one may notice that the uncertainties in $\xi$, which we denote as $\sigma_\xi$, 
is dependent on the fiducial value of $\xi$.
This is because, in the absence of  the BBN relation, there is no difference between neutrinos and anti-neutrinos 
in their effects both on the CMB and 21~cm power spectra. 
Therefore these power spectra are even functions of $\xi$, as can be also read from Eqs. \eqref{eq:rho2}-\eqref{eq:p2}
and \eqref{eq:dlnfdlny}, which respectively govern effects on the background and perturbation evolutions.
In particular for small $\xi\ll1$, these power spectra should respond linearly to $\xi^2$.
This leads that  $\sigma_\xi$ is proportional to the inverse of the fiducial $\xi$, 
while the error $\sigma_{\xi^2}\propto \xi\,\sigma_\xi$ is almost independent of the fiducial $\xi$, which 
is confirmed from Table \ref{tab:xi}, where we summarized constraints on $\xi$ for various setups  (e.g. without the BBN relation)
and fiducial values of $\xi$ for cases of $\Delta N_\nu=0$. 

Although $\sigma_\xi$ is dependent on fiducial $\xi$, we can still see that $\xi=-0.1$, which is roughly the current lower bound from 
the primordial light elements, can be detected marginally by CMBPol+SKA and significantly by CMBPol+Omniscope.
This is remarkable as this indicates that even without assuming the BBN relation, 
we may be able to obtain a constraint on $\xi$ better than one from the primordial light elements.

On the other hand, from the above figure, one may think 21~cm alone is powerful enough to give similar constraints on $\xi$ 
as those from CMB+21~cm. However, this is not true.
This can be understood by seeing that  provided a very precise observation of 21~cm, e.g., Omniscope, 
its combinations with Planck and CMBPol still differ non-negligibly.
This is due to that some cosmological parameters which degenerate with $\xi$ when only a 21~cm observation 
is adopted can be determined well by CMB. 

Let us next see the cases with the BBN relation $Y_p(\Omega_bh^2,~\xi,~\Delta N_\nu)$, though 
we here still assume $\Delta N_\nu$ to vanish. 
In this case, $\xi$ affects CMB and 21~cm observations also through $Y_p$ 
in addition to the effects we have taken into account in the case of fixed $Y_p$.
Regarding effects of $\xi$ on the CMB power spectrum, 
this indirect effect through the BBN relation is more significant than direct ones.
This can be noticed in Fig.~\ref{fig:xi_mnu_noBBN}, 
where the contours of constraints from CMB alone can be squeezed in the direction of $\xi$ 
by an order of magnitude with the BBN relation.

Fig.~\ref{fig:xi_mnu_wBBN} shows the same constraints 
as in Fig.~\ref{fig:xi_mnu_noBBN} except that the BBN relation is now taken into account in any combinations 
of observations. Compared with the previous figure, improvements brought about 
by the combination of 21~cm observations are not as dramatic as in the cases without the BBN relation.
This indirectly suggests that 21cm observations are not as sensitive to $Y_p$ as CMB.
However, the combination with SKA can reduce the size of error in $\xi$ by a few times 
from Planck alone and a similar level of improvement can be brought about by Omniscope compared to CMBPol alone.
We note that with the BBN relation being assumed, a combination of CMB and 21~cm observations
can constrain the lepton asymmetry substantially better than the primordial abundances of light elements.

Different from the cases without the BBN relation, 
one can notice that the sizes of errors in $\xi$ little depend on fiducial $\xi$ with the BBN relation. 
This is because prediction of BBN is sensitive to the sign of $\xi$. Therefore $Y_p$ responses linearly 
to $\xi$ at the lowest order. 
In particular, the most significant effect of $\xi$ on $Y_p$ is that $\xi_e$ changes the ratio of 
neutron number density to proton one when BBN starts.
Positive (negative) $\xi$ effectively boosts (suppresses) $n\to p$ 
conversion and reduces (increases) $Y_p$. 
Such an effect can break the degeneracy between $\xi$ and $-\xi$ existing without the BBN relation.



Constraints on cosmological parameters are summarized in Tables~\ref{tab:fixedYp}, \ref{tab:withBBN} and \ref{tab:variedYp}, 
where we fixed $Y_p$ to 0.25, assumed the BBN relation and varied $Y_p$ as a free parameter, respectively.
In these tables, we present constraints only for the fiducial $\xi=0.05$, as we found that 
dependencies of errors on the fiducial $\xi$ is not significant except for $\sigma_\xi$; 
as long as one considers a fiducial $\xi\le0.1$, 
errors of cosmological parameters differ by no more than 25\%. 
The only exception is $\sigma_ \xi$ which has been shown to depend on fiducial $\xi$ in the absence of the BBN relation.
Table~\ref{tab:xi} summarizes the dependence of $\sigma_ \xi$ on fiducial values of $\xi$.
Except for the cases with the BBN relation, we see that $\sigma_ \xi$ scales almost 
proportionally to the inverse of fiducial $\xi$.

\begin{table}
\begin{center}
\begin{tabular}{l|cccc}
\hline \hline \\
& $\Omega_{m}h^{2}$ &  $\Omega_{b}h^{2}$ &  $\Omega_{\Lambda}$ & $n_s $ \\ 
\hline
Planck & $ 2.86 \times 10^{  -3} $ & $ 1.95 \times 10^{  -4} $ & $ 2.01 \times 10^{  -2} $ & $ 6.06 \times 10^{  -3} $ \\
\quad + SKA phase1 & $ 3.40 \times 10^{  -4} $ & $ 7.63 \times 10^{  -5} $ & $ 2.33 \times 10^{  -3} $ & $ 2.03 \times 10^{  -3} $ \\
\quad + SKA phase2 & $ 2.52 \times 10^{  -4} $ & $ 7.40 \times 10^{  -5} $ & $ 9.26 \times 10^{  -4} $ & $ 1.42 \times 10^{  -3} $ \\
\quad + Omniscope & $ 8.16 \times 10^{  -5} $ & $ 2.42 \times 10^{  -5} $ & $ 4.18 \times 10^{  -4} $ & $ 4.81 \times 10^{  -4} $\\ 
\hline
CMBPol & $ 1.16 \times 10^{  -3} $ & $ 3.78 \times 10^{  -5} $ & $ 7.48 \times 10^{  -3} $ & $ 1.75 \times 10^{  -3} $ \\
\quad + SKA phase1 & $ 3.11 \times 10^{  -4} $ & $ 2.91 \times 10^{  -5} $ & $ 2.14 \times 10^{  -3} $ & $ 1.20 \times 10^{  -3} $ \\
\quad + SKA phase2 & $ 2.12 \times 10^{  -4} $ & $ 2.74 \times 10^{  -5} $ & $ 9.06 \times 10^{  -4} $ & $ 9.16 \times 10^{  -4} $ \\
\quad + Omniscope & $ 5.13 \times 10^{  -5} $ & $ 1.31 \times 10^{  -5} $ & $ 4.09 \times 10^{  -4} $ & $ 3.68 \times 10^{  -4} $ \\ 
\hline \\
&  $A_{s} \times 10^{10} $ & $\tau_{\rm reion}$  & $\Sigma m_{\nu}$ & $\xi$  \\ \hline
Planck & $ 2.31 \times 10^{  -1} $ & $ 4.58 \times 10^{  -3} $ & $ 1.23 \times 10^{  -1} $ & $ 9.99 \times 10^{  -1} $ \\
\quad + SKA phase1 & $ 1.88 \times 10^{  -1} $ & $ 4.36 \times 10^{  -3} $ & $ 3.69 \times 10^{  -2} $ & $ 1.58 \times 10^{  -1} $ \\
\quad + SKA phase2 & $ 1.87 \times 10^{  -1} $ & $ 4.28 \times 10^{  -3} $ & $ 2.86 \times 10^{  -2} $ & $ 1.45 \times 10^{  -1} $ \\
\quad + Omniscope & $ 1.84 \times 10^{  -1} $ & $ 4.15 \times 10^{  -3} $ & $ 1.13 \times 10^{  -2} $ & $ 6.09 \times 10^{  -2} $ \\ 
\hline
CMBPol & $ 1.10 \times 10^{  -1} $ & $ 2.46 \times 10^{  -3} $ & $ 4.26 \times 10^{  -2} $ & $ 1.51 \times 10^{  -1} $ \\
\quad + SKA phase1 & $ 1.01 \times 10^{  -1} $ & $ 2.41 \times 10^{  -3} $ & $ 1.56 \times 10^{  -2} $ & $ 8.15 \times 10^{  -2} $ \\
\quad + SKA phase2 & $ 9.95 \times 10^{  -2} $ & $ 2.37 \times 10^{  -3} $ & $ 1.10 \times 10^{  -2} $ & $ 7.69 \times 10^{  -2} $ \\
\quad + Omniscope & $ 7.81 \times 10^{  -2} $ & $ 1.78 \times 10^{  -3} $ & $ 7.15 \times 10^{  -3} $ & $ 3.19 \times 10^{  -2} $ \\ 
\hline\hline
\end{tabular} 
\caption{1$\sigma$ errors on cosmological parameters for fiducial $\xi=0.05$ for the cases with fixed $Y_p=0.25$.}
\label{tab:fixedYp}
\end{center}
\end{table}

\begin{table}
\begin{center}
\begin{tabular}{l|cccc}
\hline \hline \\
&  $\Omega_{m}h^{2}$ &  $\Omega_{b}h^{2}$ &  $\Omega_{\Lambda}$ & $ n_s $ \\ 
\hline
Planck & $ 2.41 \times 10^{  -3} $ & $ 2.13 \times 10^{  -4} $ & $ 2.09 \times 10^{  -2} $ & $ 7.06 \times 10^{  -3} $ \\
\quad + SKA phase1 & $ 3.04 \times 10^{  -4} $ & $ 9.35 \times 10^{  -5} $ & $ 2.30 \times 10^{  -3} $ & $ 2.22 \times 10^{  -3} $ \\
\quad + SKA phase2 & $ 2.02 \times 10^{  -4} $ & $ 8.64 \times 10^{  -5} $ & $ 9.21 \times 10^{  -4} $ & $ 1.44 \times 10^{  -3} $ \\
\quad + Omniscope & $ 7.94 \times 10^{  -5} $ & $ 1.54 \times 10^{  -5} $ & $ 4.15 \times 10^{  -4} $ & $ 3.54 \times 10^{  -4} $ \\ 
\hline
CMBPol & $ 9.27 \times 10^{  -4} $ & $ 4.83 \times 10^{  -5} $ & $ 7.16 \times 10^{  -3} $ & $ 2.54 \times 10^{  -3} $ \\
\quad + SKA phase1 & $ 2.75 \times 10^{  -4} $ & $ 4.16 \times 10^{  -5} $ & $ 2.11 \times 10^{  -3} $ & $ 1.46 \times 10^{  -3} $ \\
\quad + SKA phase2 & $ 1.43 \times 10^{  -4} $ & $ 4.05 \times 10^{  -5} $ & $ 9.00 \times 10^{  -4} $ & $ 1.04 \times 10^{  -3} $ \\
\quad + Omniscope & $ 4.81 \times 10^{  -5} $ & $ 1.24 \times 10^{  -5} $ & $ 4.08 \times 10^{  -4} $ & $ 3.17 \times 10^{  -4} $ \\ 
\hline \\
&  $A_{s} \times 10^{10} $ & $\tau_{\rm reion}$  & $\Sigma m_{\nu}$ & $\xi$ \\ 
\hline
Planck & $ 2.07 \times 10^{  -1} $ & $ 4.64 \times 10^{  -3} $ & $ 1.28 \times 10^{  -1} $ & $ 4.50 \times 10^{  -2} $ \\
\quad + SKA phase1 & $ 1.92 \times 10^{  -1} $ & $ 4.31 \times 10^{  -3} $ & $ 3.34 \times 10^{  -2} $ & $ 2.10 \times 10^{  -2} $ \\
\quad + SKA phase2 & $ 1.89 \times 10^{  -1} $ & $ 4.25 \times 10^{  -3} $ & $ 2.45 \times 10^{  -2} $ & $ 1.83 \times 10^{  -2} $ \\
\quad + Omniscope & $ 1.85 \times 10^{  -1} $ & $ 4.14 \times 10^{  -3} $ & $ 8.08 \times 10^{  -3} $ & $ 1.28 \times 10^{  -2} $ \\ 
\hline
CMBPol & $ 1.07 \times 10^{  -1} $ & $ 2.48 \times 10^{  -3} $ & $ 3.92 \times 10^{  -2} $ & $ 1.03 \times 10^{  -2} $ \\
\quad + SKA phase1 & $ 1.01 \times 10^{  -1} $ & $ 2.39 \times 10^{  -3} $ & $ 1.55 \times 10^{  -2} $ & $ 7.85 \times 10^{  -3} $ \\
\quad + SKA phase2 & $ 9.78 \times 10^{  -2} $ & $ 2.33 \times 10^{  -3} $ & $ 1.07 \times 10^{  -2} $ & $ 6.95 \times 10^{  -3} $ \\
\quad + Omniscope & $ 6.86 \times 10^{  -2} $ & $ 1.56 \times 10^{  -3} $ & $ 5.30 \times 10^{  -3} $ & $ 4.04 \times 10^{  -3} $ \\ 
\hline\hline
\end{tabular} 
\caption{Same as in Table~\ref{tab:fixedYp} but for the cases with the BBN relation.}
\label{tab:withBBN}
\end{center}
\end{table}

\begin{table}
\begin{center}
\begin{tabular}{l|ccccc}
\hline \hline \\
&  $\Omega_{m}h^{2}$ &  $\Omega_{b}h^{2}$ &  $\Omega_{\Lambda}$ & $ n_s $ \\ 
\hline
Planck & $ 3.31 \times 10^{  -3} $ & $ 2.27 \times 10^{  -4} $ & $ 2.11 \times 10^{  -2} $ & $ 7.56 \times 10^{  -3} $ \\
\quad + SKA phase1 & $ 3.46 \times 10^{  -4} $ & $ 1.09 \times 10^{  -4} $ & $ 2.34 \times 10^{  -3} $ & $ 2.25 \times 10^{  -3} $\\
\quad + SKA phase2 & $ 2.66 \times 10^{  -4} $ & $ 1.05 \times 10^{  -4} $ & $ 9.26 \times 10^{  -4} $ & $ 1.46 \times 10^{  -3} $ \\
\quad + Omniscope & $ 8.31 \times 10^{  -5} $ & $ 3.88 \times 10^{  -5} $ & $ 4.18 \times 10^{  -4} $ & $ 4.87 \times 10^{  -4} $ \\ 
\hline
CMBPol & $ 1.29 \times 10^{  -3} $ & $ 4.90 \times 10^{  -5} $ & $ 8.03 \times 10^{  -3} $ & $ 2.72 \times 10^{  -3} $ \\
\quad + SKA phase1 & $ 3.17 \times 10^{  -4} $ & $ 4.29 \times 10^{  -5} $ & $ 2.14 \times 10^{  -3} $ & $ 1.49 \times 10^{  -3} $  \\
\quad + SKA phase2 & $ 2.23 \times 10^{  -4} $ & $ 4.20 \times 10^{  -5} $ & $ 9.06 \times 10^{  -4} $ & $ 1.05 \times 10^{  -3} $ \\
\quad + Omniscope & $ 5.27 \times 10^{  -5} $ & $ 2.28 \times 10^{  -5} $ & $ 4.10 \times 10^{  -4} $ & $ 3.69 \times 10^{  -4} $  \\ 
\hline \\
&  $A_{s} \times 10^{10} $ & $\tau_{\rm reion}$  & $\Sigma m_{\nu}$ & $\xi$ & $Y_{p}$ \\ \hline
Planck & $ 2.32 \times 10^{  -1} $ & $ 4.66 \times 10^{  -3} $ & $ 1.28 \times 10^{  -1} $ & $ 1.12 $ & $ 1.13 \times 10^{  -2} $ \\
\quad + SKA phase1 & $ 1.92 \times 10^{  -1} $ & $ 4.36 \times 10^{  -3} $ & $ 3.70 \times 10^{  -2} $ & $ 2.10 \times 10^{  -1} $ & $ 5.90 \times 10^{  -3} $ \\
\quad + SKA phase2 & $ 1.89 \times 10^{  -1} $ & $ 4.29 \times 10^{  -3} $ & $ 2.88 \times 10^{  -2} $ & $ 2.05 \times 10^{  -1} $ & $ 5.41 \times 10^{  -3} $ \\
\quad + Omniscope & $ 1.85 \times 10^{  -1} $ & $ 4.17 \times 10^{  -3} $ & $ 1.16 \times 10^{  -2} $ & $ 8.99 \times 10^{  -2} $ & $ 3.83 \times 10^{  -3} $ \\ 
\hline
CMBPol & $ 1.10 \times 10^{  -1} $ & $ 2.49 \times 10^{  -3} $ & $ 4.47 \times 10^{  -2} $ & $ 1.85 \times 10^{  -1} $ & $ 2.83 \times 10^{  -3} $ \\
\quad + SKA phase1 & $ 1.02 \times 10^{  -1} $ & $ 2.42 \times 10^{  -3} $ & $ 1.57 \times 10^{  -2} $ & $ 1.01 \times 10^{  -1} $ & $ 2.15 \times 10^{  -3} $ \\
\quad + SKA phase2 & $ 1.00 \times 10^{  -1} $ & $ 2.37 \times 10^{  -3} $ & $ 1.11 \times 10^{  -2} $ & $ 9.89 \times 10^{  -2} $ & $ 1.96 \times 10^{  -3} $ \\
\quad + Omniscope & $ 7.94 \times 10^{  -2} $ & $ 1.91 \times 10^{  -3} $ & $ 7.47 \times 10^{  -3} $ & $ 4.93 \times 10^{  -2} $ & $ 1.31 \times 10^{  -3} $  \\ 
\hline\hline
\end{tabular} 
\caption{Same as in Table~\ref{tab:fixedYp} but for the cases with freely varying $Y_p$.}
\label{tab:variedYp}
\end{center}
\end{table}

\begin{table}
\begin{itemize}
\item Fixing $Y_p=0.25$ 
\begin{center}
\begin{tabular}{l|ccc}
\hline \hline 
& $\xi=-0.1$ & $\xi=0.05$ & $\xi=0.01$ \\
\hline
Planck & $5.01\times 10^{-1}$ & $ 9.99 \times 10^{  -1} $ & $4.88$ \\
\quad + SKA phase1 & $ 7.85 \times 10^{  -2} $ & $ 1.58 \times 10^{  -1} $ & $ 7.73 \times 10^{  -1} $ \\
\quad + SKA phase1  & $ 7.23 \times 10^{  -2} $ & $ 1.45 \times 10^{  -1} $ & $ 6.76 \times 10^{  -1} $ \\
\quad + Omniscope  & $ 3.02 \times 10^{  -2} $  & $ 6.09 \times 10^{  -2} $ & $ 2.62 \times 10^{  -1} $ \\
\hline
CMBPol & $ 7.55 \times 10^{  -2} $ & $ 1.51 \times 10^{  -1} $ & $ 7.50 \times 10^{  -1} $ \\
\quad + SKA phase1 & $ 4.07 \times 10^{  -2} $ & $ 8.15 \times 10^{  -2} $ & $ 4.05 \times 10^{  -1} $ \\
\quad + SKA phase2 & $ 3.84 \times 10^{  -2} $ & $ 7.69 \times 10^{  -2} $ & $ 3.76 \times 10^{  -1} $ \\
\quad + Omniscope & $ 1.59 \times 10^{  -2} $ & $ 3.19 \times 10^{  -2} $ & $ 1.52 \times 10^{  -1} $ \\
\hline\hline
\end{tabular}
\end{center}
\item With the BBN relation 
\begin{center}
\begin{tabular}{l|ccc}
\hline\hline
& $\xi=-0.1$ & $\xi=0.05$ & $\xi=0.01$ \\
\hline
Planck & $3.72 \times 10^{-2}$ & $ 4.50 \times 10^{  -2} $ & $ 4.29\times 10^{-2}$ \\
\quad + SKA phase1 & $ 1.49 \times 10^{  -2} $ & $ 2.10 \times 10^{  -2} $ & $ 1.90 \times 10^{  -2} $ \\
\quad + SKA phase2 & $ 1.29 \times 10^{  -2} $ & $ 1.83 \times 10^{  -2} $ & $ 1.65 \times 10^{  -2} $ \\
\quad + Omniscope & $ 7.66 \times 10^{  -3} $ & $ 1.28 \times 10^{  -2} $ & $ 1.10 \times 10^{  -2} $ \\
\hline
CMBPol & $ 7.82 \times 10^{  -3} $ & $ 1.03 \times 10^{  -2} $ & $ 9.68 \times 10^{  -3} $ \\
\quad + SKA phase1 & $ 5.89 \times 10^{  -3} $ & $ 7.85 \times 10^{  -3} $ & $ 7.31 \times 10^{  -3} $ \\
\quad + SKA phase2 & $ 5.25 \times 10^{  -3} $ & $ 6.95 \times 10^{  -3} $ & $ 6.47 \times 10^{  -3} $ \\
\quad + Omniscope & $ 2.86 \times 10^{  -3} $ & $ 4.04 \times 10^{  -3} $ & $ 3.65 \times 10^{  -3} $ \\
\hline\hline
\end{tabular}
\end{center}
\item Freely varying $Y_p$
\begin{center}
\begin{tabular}{l|ccc}
\hline\hline
& $\xi=-0.1$ & $\xi=0.05$ & $\xi=0.01$ \\
\hline
Planck & $ 5.61 \times 10^{  -1} $ & $ 1.12 $ & $ 5.42 $ \\
\quad + SKA phase1 & $ 1.05 \times 10^{  -1} $ & $ 2.10 \times 10^{  -1} $ & $ 1.02 $ \\
\quad + SKA phase2 & $ 1.02 \times 10^{  -1} $ & $ 2.05 \times 10^{  -1} $ & $ 9.06 \times 10^{  -1} $ \\
\quad + Omniscope & $ 4.48 \times 10^{  -2} $ & $ 8.99 \times 10^{  -2} $ & $ 3.39 \times 10^{  -1} $ \\
\hline
CMBPol & $ 9.24 \times 10^{  -2} $ & $ 1.85 \times 10^{  -1} $ & $ 9.17 \times 10^{  -1} $\\
\quad + SKA phase1 & $ 5.07 \times 10^{  -2} $ & $ 1.01 \times 10^{  -1} $ & $ 5.03 \times 10^{  -1} $ \\
\quad + SKA phase2 & $ 4.95 \times 10^{  -2} $ & $ 9.89 \times 10^{  -2} $ & $ 4.79 \times 10^{  -1} $ \\
\quad + Omniscope & $ 2.46 \times 10^{  -2} $ & $ 4.93 \times 10^{  -2} $ & $ 2.24 \times 10^{  -1} $ \\
\hline\hline
\end{tabular}
\end{center}
\end{itemize}
\caption{Dependence of $\sigma_\xi$ on the fiducial value of $\xi$.}
\label{tab:xi}
\end{table}

%
%

%
%

\subsection{Cases with extra radiation}
So far we have been investigating constraints on $\xi$ in combination with CMB and 21~cm observations.
Having observed that the combination of observations can improve constraints on $\xi$ from only CMB ones, 
we extend our analysis to consider cosmological models with not only nonzero $\xi$ but also 
extra (dark) radiation other than active neutrinos. 
%
Throughout this subsection, we assume that the extra radiation is massless. 
In addition, we assume the BBN relation  $Y_p(\Omega_bh^2,~\xi,~\Delta N_\nu)$, 
which allows us to distinguish $\xi$ and $\Delta N_\nu$ even if the active neutrinos are almost massless.



\begin{figure}[htbp]
  \begin{center}
    \resizebox{120mm}{!}{
     \includegraphics{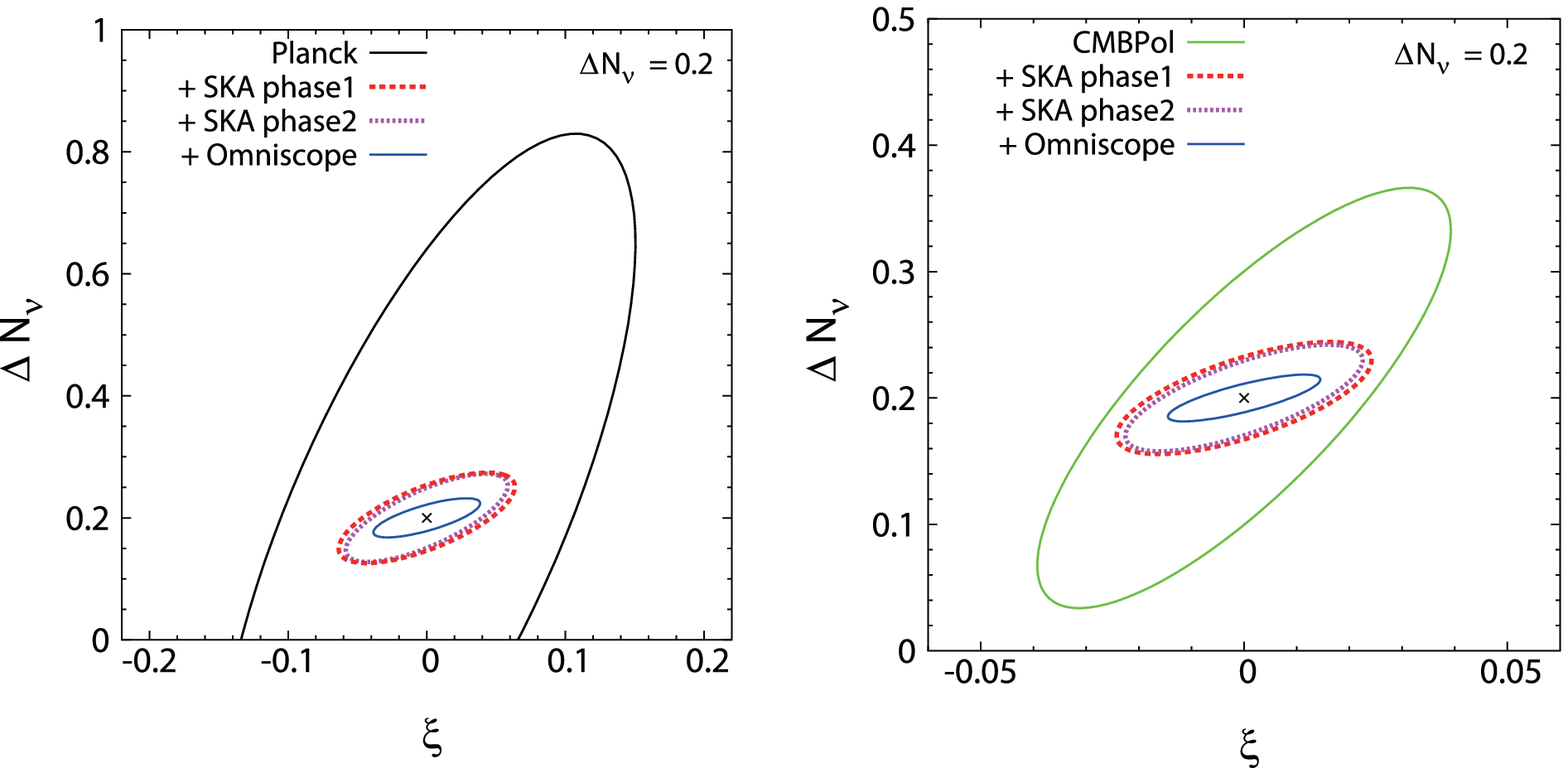}}
    \resizebox{120mm}{!}{
     \includegraphics{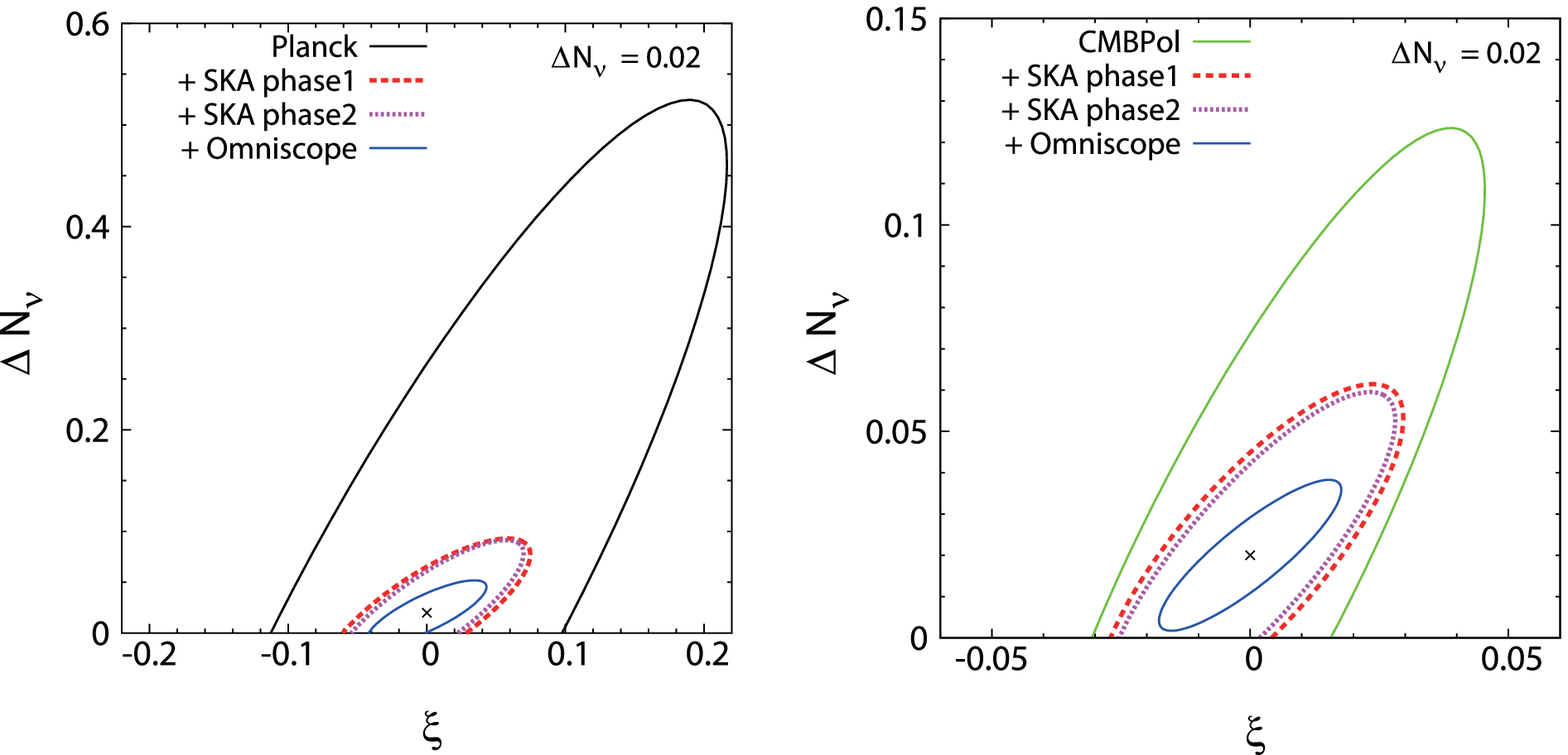}}
    \resizebox{120mm}{!}{
     \includegraphics{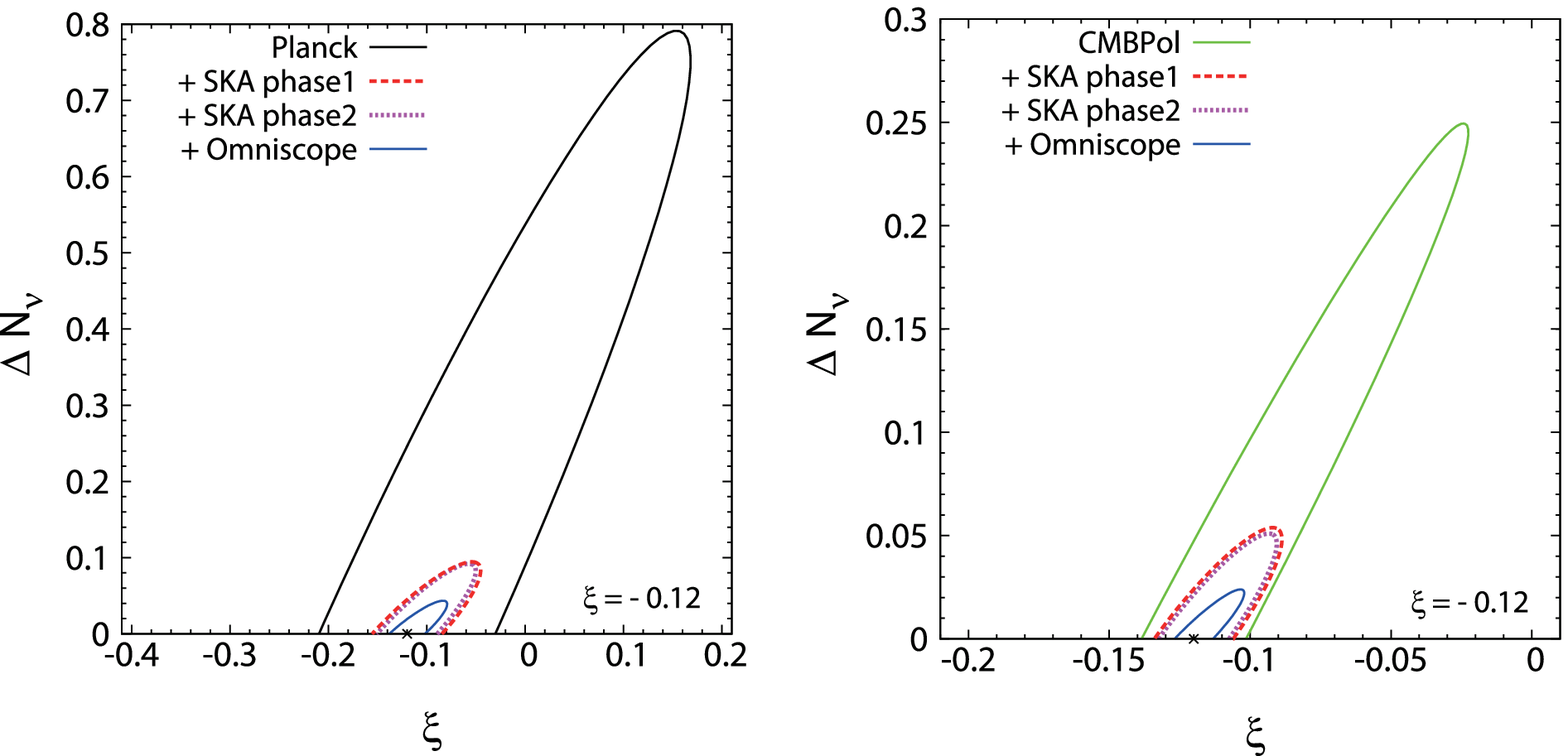}}
  \end{center}
  \caption{Expected 2$\sigma$ constraints on the $\xi$--$\Delta N_\nu$ plane. In this figure, the BBN relation is assumed.
  As fiducial values of ($\xi,~\Delta N_\nu$), we here adopt (0.2, 0), (0.02, 0) and (0, $-0.12$) in the top, middle and bottom panels, respectively.
  Note that scales differ among different panels.}
  \label{fig:xi_dNnu_wBBN}
\end{figure}

In Fig.~\ref{fig:xi_dNnu_wBBN}, we plot 2$\sigma$ constraints in the $\xi$--$\Delta N_\nu$ plane from
CMB alone as well as combinations of CMB and 21~cm.  Three different fiducial models 
$(\xi, \Delta N_\nu)=(0,~0.2)$, (0,~0.02) and ($-0.12$,~0) are adopted here.
We note that the latter two fiducial models give the similar effective numbers of neutrino species
when neutrinos are relativistic.
We can see that CMB alone cannot constrain $\Delta N_\nu$ tightly. 
Moreover, the sizes of 2$\sigma$ contours in the $\Delta N_\nu$ direction are dependent on fiducial 
parameters $\xi$ and $\Delta N_\nu$.
This dependency should be suggesting that 
observations are not enough constraining and the likelihood surface in the $\xi$-$\Delta N_\nu$ plane
deviates from Gaussian cases to some extent.
This may lead that when one explores constraints in a full parameter space using the Markov chain Monte Carlo, e.g., CosmoMC~\cite{Lewis:2002ah}, 
resulting constraints would be somewhat less stringent than forecasts based on the Fisher matrix analysis.
However, once we combine 21~cm observations, the constraints on $\Delta N_\nu$ greatly improve.
Moreover, the size of errors become almost independent of the fiducial values of $\xi$ and $\Delta N_\nu$ 
by an order of magnitude.
This shows that combinations of CMB and 21~cm line observations will be promising 
to disentangle degenerating $\xi$ and $\Delta N_\nu$.
In Table~\ref{tab:xi_dNnu_wBBN}, we present the 1$\sigma$ constraints only for $\xi$ and $\Delta N_\nu$.
We note that regarding the constraints on other cosmological parameters, 
the inclusion of $\Delta N_\nu$ does not degrade most of them significantly, or, by at most 50 \%.
Only exceptions are the constants on $\Omega_mh^2$ from Planck alone and
$\Omega_bh^2$ from Planck+Omniscope and CMBPol+Omniscope, which are degraded by 2-3 times.


\begin{table}
\begin{itemize}
\item fiducial $(\xi,~\Delta N_\nu)=(0,~0.2)$
\begin{center}
\begin{tabular}{l|cc}
\hline \hline 
\\
& $\xi$ & $\Delta N_\nu$ \\
\hline 
Planck & $ 6.07 \times 10^{-  2} $ & $ 2.54 \times 10^{  -1} $ \\
\quad + SKA phase1 & $ 2.56 \times 10^{  -2} $ & $ 2.99 \times 10^{  -2} $ \\
\quad + SKA phase2  & $ 2.36 \times 10^{  -2} $ & $ 2.91 \times 10^{  -2} $ \\
\quad + Omniscope  & $ 1.55 \times 10^{  -2} $  & $ 1.29 \times 10^{  -2} $ \\
\hline
CMBPol & $ 1.58 \times 10^{-  2} $ & $ 6.71 \times 10^{  -2} $ \\
\quad + SKA phase1 & $ 9.77 \times 10^{  -3} $ & $ 1.79 \times 10^{  -2} $ \\
\quad + SKA phase2  & $ 9.09 \times 10^{  -3} $ & $ 1.70 \times 10^{  -2} $ \\
\quad + Omniscope  & $ 5.83 \times 10^{  -3} $  & $ 7.47 \times 10^{  -3} $ \\
\hline\hline
\end{tabular}
\end{center}
\item fiducial $(\xi,~\Delta N_\nu)=(0,~0.02)$
\begin{center}
\begin{tabular}{l|cc}
\hline \hline 
\\
& $\xi$ & $\Delta N_\nu$ \\
\hline 
Planck & $ 8.74 \times 10^{-  2} $ & $ 2.04 \times 10^{  -1} $ \\
\quad + SKA phase1 & $ 3.01 \times 10^{  -2} $ & $ 2.94 \times 10^{  -2} $ \\
\quad + SKA phase2  & $ 2.82 \times 10^{  -2} $ & $ 2.88 \times 10^{  -2} $ \\
\quad + Omniscope  & $ 1.74 \times 10^{  -2} $  & $ 1.28 \times 10^{  -2} $ \\
\hline
CMBPol & $ 1.83 \times 10^{-  2} $ & $ 4.17 \times 10^{  -2} $ \\
\quad + SKA phase1 & $ 1.20 \times 10^{  -2} $ & $ 1.67 \times 10^{  -2} $ \\
\quad + SKA phase2  & $ 1.13 \times 10^{  -2} $ & $ 1.59 \times 10^{  -2} $ \\
\quad + Omniscope  & $ 7.11 \times 10^{  -3} $  & $ 7.37 \times 10^{  -3} $ \\
\hline\hline
\end{tabular}
\end{center}
\item fiducial $(\xi,~\Delta N_\nu)=(-0.12,~0)$
\begin{center}
\begin{tabular}{l|cc}
\hline \hline 
\\
& $\xi$ & $\Delta N_\nu$ \\
\hline 
Planck & $ 1.16 \times 10^{-  1} $ & $ 3.19 \times 10^{  -1} $ \\
\quad + SKA phase1 & $ 3.02 \times 10^{  -2} $ & $ 3.81 \times 10^{  -2} $ \\
\quad + SKA phase2  & $ 2.82 \times 10^{  -2} $ & $ 3.71 \times 10^{  -2} $ \\
\quad + Omniscope  & $ 1.64 \times 10^{  -2} $  & $ 1.75 \times 10^{  -2} $ \\
\hline
CMBPol & $ 3.93 \times 10^{-  2} $ & $ 1.01 \times 10^{  -1} $ \\
\quad + SKA phase1 & $ 1.26 \times 10^{  -2} $ & $ 2.17 \times 10^{  -2} $ \\
\quad + SKA phase2  & $ 1.19 \times 10^{  -2} $ & $ 2.06 \times 10^{  -2} $ \\
\quad + Omniscope  & $ 7.22 \times 10^{  -3} $  & $ 9.65 \times 10^{  -3} $ \\
\hline\hline
\end{tabular}
\end{center}
\end{itemize}
\caption{1 $\sigma$ errors on $\xi$ and $\Delta N_\nu$ for the case with the BBN relation 
and their dependence on fiducial $(\xi,~\Delta N_\nu)$}
\label{tab:xi_dNnu_wBBN}
\end{table}


\section{Summary}

We have conducted a forecast for constraints on the lepton asymmetry $\xi$
from the future 21~cm observations.
A detection of a finite $\xi$ from cosmological observations can give unique 
implications for the origin of the baryon asymmetry in our Universe.
In our analysis, we have adopted the power spectra of the 
21~cm signal from redshifts before the reionization, in combination
with those of CMB.
When we consider constraints on $\xi$ in the absence extra radiation, 
we have found that, even without assuming the BBN relation, 
combinations of 21~cm and CMB observations can
constrain $\xi$ with a better accuracy than the primordial 
abundances of light elements, which cannot be achieved by CMB alone.
On the other hand, once the BBN relation has been taken into account, 
even the sensitivity of CMB observations alone to $\xi$ substantially improves, 
21~cm observations however can still improve the constraints 
and be useful in constraining the lepton asymmetry.
In addition, we have also investigated constraints on $\xi$ in the presence of 
some extra radiation. We have shown that 21~cm observations
can substantially improve the constraints on $\Delta N_\nu$ from CMB alone,
and allow us to distinguish between the lepton asymmetry and 
extra radiation.
Our results should be indicating that 21~cm observations can be
a powerful probe of neutrinos and the origin of matter in the Universe.

%
%
%
%
%

\section*{Acknowledgments}

This work is partially supported by the Grant-in-Aid for Scientific
research from the Ministry of Education, Science, Sports, and Culture,
Japan, Nos. 21111006, 22244030, 23540327, 26105520 (K.K.), 23.5622
(T.S.), 25.4260 (Y.O.) and 23740195 (T.T.). K.K. is supported by the
Center for the Promotion of Integrated Science (CPIS) of Sokendai
(1HB5804100).  T.S. is supported by the Academy of Finland grant
1263714.

\appendix
\bigskip
\bigskip
\noindent 
{\Large \bf Appendix}

\section{Expressions for the coefficients $C_i$}
\label{sec:appA}
Below we give explicit expressions for the coefficients $C_i$, which are necessary to 
obtain Eqs.~\eqref{eq:rho_nr} and \eqref{eq:p_nr}. 
\begin{eqnarray}
C_0 & = & \frac{2}{e^y+1}, \\
C_2 & = & \frac{e^y \left(e^y-1\right)}{\left(e^y+1\right)^3}, \\
C_4 & = & \frac{e^y \left(11 e^y-11 e^{2 y}+e^{3 y}-1\right)}{12 \left(e^y+1\right)^5}, \\
C_6 & = &  \frac{e^y \left(57 e^y-302 e^{2 y}+302 e^{3 y}-57 e^{4 y}+e^{5 y}-1\right)}{360
   \left(e^y+1\right)^7}, \\ 
C_8 & = &  \frac{e^y \left(247 e^y-4293 e^{2 y}+15619 e^{3 y}-15619 e^{4 y}+4293 e^{5
   y}-247 e^{6 y}+e^{7 y}-1\right)}{20160 \left(e^y+1\right)^9},\\
C_{10} & = &   \frac{e^y \! \left(1013 e^y\!-\!47840 e^{2 y}\!+\!455192 e^{3 y}\!-\!1310354 e^{4 y}\!+\!1310354
   e^{5 y}\!-\!455192 e^{6 y}\right)}
   {1814400\left(e^y+1\right)^{11}} \nonumber \\
   && + \frac{e^y \! \left(47840 e^{7 y}\!-\!1013 e^{8 y}\!+\!e^{9 y}\!-\!1\right)}
   {1814400\left(e^y+1\right)^{11}}.
\end{eqnarray}

\section{Non-relativistic limit of $\rho_{\nu}+\rho_{\bar{\nu}}$ and $p_{\nu}+p_{\bar{\nu}}$
for any $\xi$}
\label{sec:appB}

Below we show the exact solutions for the $\rho_{\nu}+\rho_{\bar{\nu}}$ and $p_{\nu}+p_{\bar{\nu}}$
for any $\xi$ in non-relativistic limit by using polylogarithm ${\rm Li}_{s}(z)$, which is one of special functions.
\begin{eqnarray}
\!\!\!\!\!\!\!\!\!\!
\rho_\nu \!+\! \rho_{\bar{\nu}} \!
&\simeq &\!
\frac{T_{\nu }^4 a \tilde{m}}{2\pi^2 } \int_0^{\infty} y^2 dy  
\left[ 1 + \frac12 \left( \frac{y}{a \tilde{m}} \right)^2  \right]
\left( \frac{1}{e^{y + \xi} +1}  +  \frac{1}{e^{y - \xi} +1} \right) \notag \\
&=&\!
\frac{T_{\nu }^4 a \tilde{m}}{2\pi^2 }\!
\left[
\!-2\!\left\{\!{\rm Li}_{3}(\!-e^{-\xi})+{\rm Li}_{3}(\!-e^{\xi})\!\right\}\!
\right] 
\! +\! \frac{T_{\nu }^4}{4\pi^2  a \tilde{m}}\! \!
\left[
\!-24\!\left\{\!{\rm Li}_{5}(\!-e^{-\xi})+{\rm Li}_{5}(\!-e^{\xi})\!\right\}\!
\right]\!, \\
\!\!\!\!\!\!\!\!\!\!
p_{\nu}\!+\!p_{\bar{\nu}}\!
&\simeq&\!
\frac{T_{\nu }^4 }{6\pi^2 a \tilde{m}} \int_0^{\infty} y^4 dy  
\left[ 1 - \frac12 \left( \frac{y}{a \tilde{m}} \right)^2  \right]
\left( \frac{1}{e^{y + \xi} +1}  +  \frac{1}{e^{y - \xi} +1} \right) \notag \\
&=&\!
\frac{T_{\nu }^4 }{6\pi^2 a \tilde{m}} \!\!
\left[
\!-24\!\left\{\!{\rm Li}_{5}(\!-e^{-\xi})\!+\!{\rm Li}_{5}(\!-e^{\xi})\!\right\}\!
\right]
\! - \! \frac{T_{\nu }^4}{12\pi^2  (a \tilde{m})^{3}}\! \!
\left[
\!-720\!\left\{\!{\rm Li}_{7}(\!-e^{-\xi})\!+\!{\rm Li}_{7}(\!-e^{\xi})\!\right\}\!
\right]\!.
\end{eqnarray} 
If we expand these formulas around  $\xi = 0$ ,
they reduce to Eqs.~\eqref{eq:rho_nr} and \eqref{eq:p_nr}.

\section{BBN relation}
\label{sec:BBNrelation}

In the early universe with a higher temperature than {\cal O}(1)~MeV
the inter-converting reactions between
neutron and proton through the weak interaction ($n+ e^+
\leftrightarrow p + \nu_e$, $n+ \bar{\nu}_e \leftrightarrow p + e^-$,
and $n \leftrightarrow p + e^- + \nu_e$) were sufficiently rapid. In
this case the neutron to proton ratio obeys its thermal equilibrium
value,
\begin{eqnarray}
  \label{eq:n2pth}
 \frac{n}{p} = \exp\left[ - \frac{\Delta m_{np} + \mu_{\nu_e}}{T}\right] = \exp\left[ -\frac{\Delta m_{np}}{T} - \xi_{\nu_e}\right],
\end{eqnarray}
with the mass difference $\Delta m_{np} = 1.3$~MeV. Here we explicitly
wrote the degeneracy parameter of $\nu_e$ to be $\xi_{\nu_e} =
\mu_{\nu_e}/T_{\nu}$ with $\mu_{\nu_e}$ being the chemical potential of
$\nu_e$. It is remarkable that the electron's chemical potential $\mu_{e^{-}}$ must
be much smaller than that of $\nu_e$ because of the neutrality of the
Universe $\xi_{e} = \mu_{e^{-}}/T \sim O(\eta) \ll \xi_{\nu_e}$ 
with  $T$ and $\eta$ being the photon temperature and the baryon-to-photon ratio, respectively.
Accordingly
$\xi_{\nu_e}$ affects the freezeout value of $n/p$, which can change
the light element abundances. 
In particular $Y_p$ depends on
$\xi_{\nu_e}$ in addition to $\eta$ (or $\Omega_bh^2$) and $N_\nu$.
Thus $Y_p$ is related to those three parameters 
i.e., $Y_p$=$Y_p(\Omega_bh^2,~\xi_{\nu_e},~\Delta N_\nu)$,
which is called "the BBN relation."

Since we need quite a precise value of $Y_p$ in the current studies,
we numerically compute $Y_p$ as functions of those three parameters
without adopting known fitting formula (e.g., given in
Ref.~\cite{Steigman:2007xt}). In this computation, we have used the
most recent data for nuclear reaction
rates~\cite{Smith:1992yy,Angulo:1999zz,Cyburt:2001pp,Serpico:2004gx, Cyburt:2008up}.

\begin{figure}[htbp]
\begin{center}
\resizebox{120mm}{!}{
\includegraphics{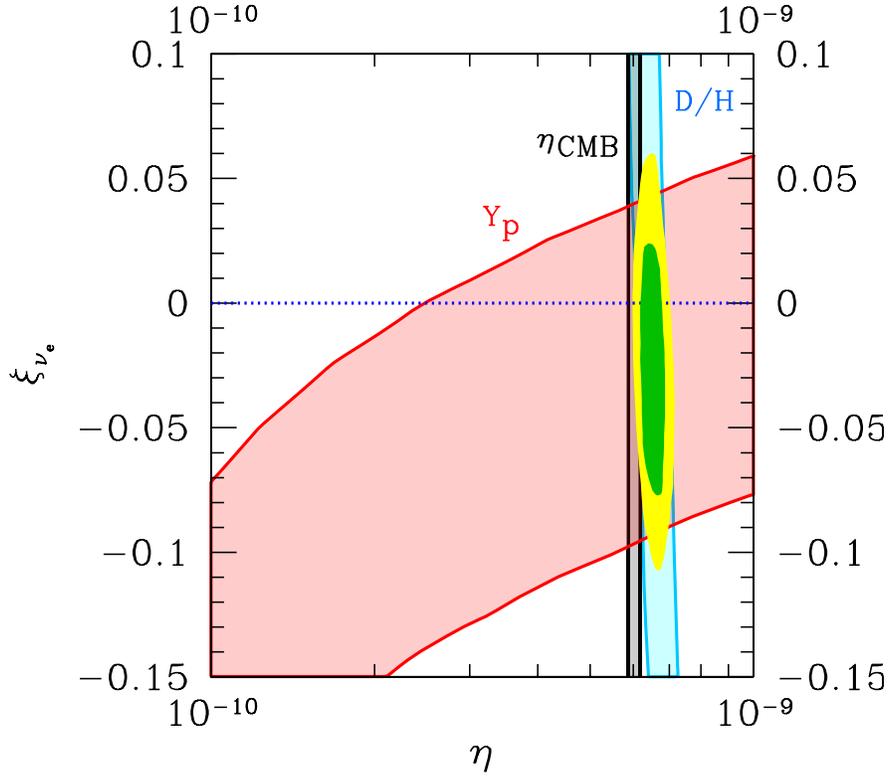}
}
\end{center}
\caption{Regions allowed by the BBN alone in the
  $\eta-\xi_{\nu_e}$ plane. The 68$\%$ and the 95$\%$
  C.L. contours are plotted, respectively.  Here we set $\Delta
  N_\nu=0$. 
The vertical band represents the baryon to photon ratio reported by Planck
$\eta = (6.04 \pm 0.15)\times10^{-10}$ at 95$\%$C.L..
The line of each light element corresponds to the individual constraint at 95$\%$ C.L..
}
  \label{fig:etaXi2014}
\end{figure}

In Fig.~\ref{fig:etaXi2014}, as a reference, we plotted allowed
regions in the $\eta-\xi_{\nu_e}$plane at the 68$\%$ and the
95$\%$ C.L, respectively. Here we set $\Delta N_\nu=0$. We have
adopted the following observational light element abundances, $Y_p =
0.2534 \pm 0.0083$ (68$\%$) \cite{Aver:2011bw} and D/H=
$n_{\rm  D}/n_{\rm H} = (2.535 \pm 0.050) \times 10^{-5}$
(68$\%$)~\cite{Pettini:2012ph}.

\begin{thebibliography}{100}

\bibitem{Kohri:1996ke} 
  K.~Kohri, M.~Kawasaki and K.~Sato,
  Astrophys.\ J.\  {\bf 490}, 72 (1997)
  [astro-ph/9612237].

\bibitem{Sato:1998nf} 
  K.~Sato, K.~Kohri and M.~Kawasaki,
  RESCEU-59-98.

\bibitem{Popa:2008tb} 
  L.~A.~Popa and A.~Vasile,
  JCAP {\bf 0806}, 028 (2008)
  [arXiv:0804.2971 [astro-ph]].

\bibitem{Shiraishi:2009fu} 
  M.~Shiraishi, K.~Ichikawa, K.~Ichiki, N.~Sugiyama and M.~Yamaguchi,
  JCAP {\bf 0907}, 005 (2009)
  [arXiv:0904.4396 [astro-ph.CO]].

\bibitem{Caramete:2013bua} 
  A.~Caramete and L.~A.~Popa,
  arXiv:1311.3856 [astro-ph.CO].


\bibitem{Casas:1997gx} 
  A.~Casas, W.~Y.~Cheng and G.~Gelmini,
  Nucl.\ Phys.\ B {\bf 538}, 297 (1999)
  [hep-ph/9709289].

\bibitem{MarchRussell:1999ig} 
  J.~March-Russell, H.~Murayama and A.~Riotto,
  JHEP {\bf 9911}, 015 (1999)
  [hep-ph/9908396].

\bibitem{McDonald:1999in} 
  J.~McDonald,
  Phys.\ Rev.\ Lett.\  {\bf 84}, 4798 (2000)
  [hep-ph/9908300].


\bibitem{Kawasaki:2002hq} 
  M.~Kawasaki, F.~Takahashi and M.~Yamaguchi,
  Phys.\ Rev.\ D {\bf 66}, 043516 (2002)
  [hep-ph/0205101].


\bibitem{Takahashi:2003db} 
  F.~Takahashi and M.~Yamaguchi,
  Phys.\ Rev.\ D {\bf 69}, 083506 (2004)
  [hep-ph/0308173].


\bibitem{Schwarz:2009ii} 
  D.~J.~Schwarz and M.~Stuke,
  JCAP {\bf 0911}, 025 (2009)
  [Erratum-ibid.\  {\bf 1010}, E01 (2010)]
  [arXiv:0906.3434 [hep-ph]].

\bibitem{Semikoz:2009ye} 
  V.~B.~Semikoz, D.~D.~Sokoloff and J.~W.~F.~Valle,
  Phys.\ Rev.\ D {\bf 80}, 083510 (2009)
  [arXiv:0905.3365 [hep-ph]].

\bibitem{Gordon:2003hw} 
  C.~Gordon and K.~A.~Malik,
  Phys.\ Rev.\ D {\bf 69}, 063508 (2004)
  [astro-ph/0311102].

\bibitem{Lyth:2002my} 
  D.~H.~Lyth, C.~Ungarelli and D.~Wands,
  Phys.\ Rev.\ D {\bf 67}, 023503 (2003)
  [astro-ph/0208055].

\bibitem{DiValentino:2011sv} 
  E.~Di Valentino, M.~Lattanzi, G.~Mangano, A.~Melchiorri and P.~Serpico,
  Phys.\ Rev.\ D {\bf 85}, 043511 (2012)
  [arXiv:1111.3810 [astro-ph.CO]].

\bibitem{skaweb}
{\tt http://www.skatelescope.org}


\bibitem{Tegmark:2009kv} 
  M.~Tegmark and M.~Zaldarriaga,
  Phys.\ Rev.\ D {\bf 82}, 103501 (2010)
  [arXiv:0909.0001 [astro-ph.CO]].

\bibitem{Planck:2006aa} 
  J.~Tauber {\it et al.}  [Planck Collaboration],
  astro-ph/0604069.

\bibitem{Baumann:2008aq} 
  D.~Baumann {\it et al.}  [CMBPol Study Team Collaboration],
  AIP Conf.\ Proc.\  {\bf 1141}, 10 (2009)
  [arXiv:0811.3919 [astro-ph]].


\bibitem{Ma:1995ey} 
  C.~-P.~Ma and E.~Bertschinger,
  Astrophys.\ J.\  {\bf 455}, 7 (1995)
  [astro-ph/9506072].
  
\bibitem{Lesgourgues:1999wu} 
  J.~Lesgourgues and S.~Pastor,
  Phys.\ Rev.\ D {\bf 60}, 103521 (1999)
  [hep-ph/9904411].
  
  \bibitem{Lewis:1999bs} 
  A.~Lewis, A.~Challinor and A.~Lasenby,
  Astrophys.\ J.\  {\bf 538}, 473 (2000)
  [astro-ph/9911177].
  

\bibitem{Furlanetto:2006jb} 
  S.~Furlanetto, S.~P.~Oh and F.~Briggs,
Phys.\ Rept.\  {\bf 433}, 181 (2006)
[astro-ph/0608032].  

\bibitem{Pritchard:2011xb} 
  J.~R.~Pritchard and A.~Loeb,
  Rept.\ Prog.\ Phys.\  {\bf 75}, 086901 (2012)
  [arXiv:1109.6012 [astro-ph.CO]].

\bibitem{McQuinn:2005hk} 
  M.~McQuinn, O.~Zahn, M.~Zaldarriaga, L.~Hernquist and S.~R.~Furlanetto,
  Astrophys.\ J.\  {\bf 653}, 815 (2006)
  [astro-ph/0512263].


\bibitem{Oyama:2012tq} 
  Y.~Oyama, A.~Shimizu and K.~Kohri,
  Phys.\ Lett.\ B {\bf 718}, 1186 (2013)
  [arXiv:1205.5223 [astro-ph.CO]].

\bibitem{Kohri:2013mxa} 
  K.~Kohri, Y.~Oyama, T.~Sekiguchi and T.~Takahashi,
  JCAP {\bf 1310}, 065 (2013)
  [arXiv:1303.1688 [astro-ph.CO]].



\bibitem{Dolgov:2002ab} 
  A.~D.~Dolgov, S.~H.~Hansen, S.~Pastor, S.~T.~Petcov, G.~G.~Raffelt and D.~V.~Semikoz,
  Nucl.\ Phys.\ B {\bf 632}, 363 (2002)
  [hep-ph/0201287].


\bibitem{Ade:2013zuv} 
  P.~A.~R.~Ade {\it et al.}  [Planck Collaboration],
  arXiv:1303.5076 [astro-ph.CO].

\bibitem{Steigman:2012ve} 
  G.~Steigman,
  Adv.\ High Energy Phys.\  {\bf 2012}, 268321 (2012)
  [arXiv:1208.0032 [hep-ph]].

\bibitem{Lewis:2002ah} 
  A.~Lewis and S.~Bridle,
  Phys.\ Rev.\ D {\bf 66}, 103511 (2002)
  [astro-ph/0205436].

\bibitem{Steigman:2007xt} 
  G.~Steigman,
  Ann.\ Rev.\ Nucl.\ Part.\ Sci.\  {\bf 57}, 463 (2007)
  [arXiv:0712.1100 [astro-ph]].


\bibitem{Smith:1992yy}
  M.~S.~Smith, L.~H.~Kawano and R.~A.~Malaney,
  Astrophys.\ J.\ Suppl.\  {\bf 85}, 219 (1993).


\bibitem{Angulo:1999zz}
  C.~Angulo {\it et al.},
  Nucl.\ Phys.\  A {\bf 656}, 3 (1999).


\bibitem{Cyburt:2001pp}
  R.~H.~Cyburt, B.~D.~Fields and K.~A.~Olive,
  New Astron.\  {\bf 6}, 215 (2001);
  R.~H.~Cyburt,
  Phys.\ Rev.\  D {\bf 70}, 023505 (2004).

\bibitem{Serpico:2004gx}
  P.~D.~Serpico, S.~Esposito, F.~Iocco, G.~Mangano, G.~Miele and O.~Pisanti,
  JCAP {\bf 0412}, 010 (2004).

\bibitem{Cyburt:2008up}
  R.~H.~Cyburt and B.~Davids,
  arXiv:0809.3240 [nucl-ex].

\bibitem{Aver:2011bw} 
  E.~Aver, K.~A.~Olive and E.~D.~Skillman,
  JCAP {\bf 1204}, 004 (2012)
  [arXiv:1112.3713 [astro-ph.CO]].

\bibitem{Pettini:2012ph} 
  M.~Pettini and R.~Cooke,
  Mon.\ Not.\ Roy.\ Astron.\ Soc.\  {\bf 425}, 2477 (2012)
  [arXiv:1205.3785 [astro-ph.CO]].

\end{thebibliography}
\end{document}